\documentclass[12pt]{iopart}
\usepackage{iopams}
\usepackage{graphicx}
\usepackage{dsfont}
\usepackage{pxfonts}

\newcommand{\half}{\mbox{$\textstyle \frac{1}{2}$}} 
\newcommand{\quat}{\mbox{$\textstyle \frac{1}{4}$}} 
\newcommand{\re}{{\rm e}}
\newcommand{\ri}{{\rm i}}
\newcommand{\rd}{{\rm d}}

\begin{document}

\title[Complex extension of Wigner's theorem]
{Complex extension of Wigner's theorem}

\author[Brody]{Dorje~C~Brody}

\address{Mathematical Sciences, Brunel University, Uxbridge UB8 3PH, UK}

\begin{abstract}
Wigner's theorem asserts that an isometric (probability conserving) transformation on a quantum 
state space must be generated by a Hamiltonian that is Hermitian. It is shown that when the 
Hermiticity condition on the Hamiltonian is relaxed, we obtain the following complex generalisation 
of Wigner's theorem: a holomorphically projective (complex geodesic-curves preserving) 
transformation on a quantum state space must be generated by a Hamiltonian that is not 
necessarily Hermitian. 
\end{abstract}

\submitto{\JPA}

\vskip 0.5cm

\noindent \textbf{Introduction}. 
Unitarity is perhaps the most fundamental symmetry governing the laws of nature. If asked why 
unitarity is important, most physicists will offer a version of Wigner's theorem as an 
argument in support of the requirement of unitarity. In essence, Wigner's theorem asserts that 
isometries (overlap-distance preserving maps) of the quantum state space can only be generated 
by unitary (or antiunitary) transformations \cite{wigner}. As a generator of a unitary motion, the 
Hamiltonian ${\hat H}$ of a system therefore has to be Hermitian (or self-adjoint). 

Over the past decade or so, however, there have been a lot of interests in understanding 
properties of complex Hamiltonians that are not Hermitian. Evidently, if the generator ${\hat K} 
\neq {\hat K}^\dagger$ of the dynamics is not Hermitian, then the evolution cannot give rise to 
an isometry. A natural question thus arising is: What is the symmetry property of the dynamical 
evolution generated by a Hamiltonian that is not Hermitian? The purpose of the present paper 
is to offer the solution to this question, which can be viewed as a complex generalisation of the 
Wigner theorem. 

The main findings can be summarised as follows. We consider the motion on the Hilbert space 
generated by an operator of the form $\re^{{\rm i}{\hat K}t}$ (where ${\hat K}$ is not necessarily 
Hermitian), project down the motion to the state space (the 
space of rays through the origin of the Hilbert space), and analyse symmetry properties of 
the dynamics. We shall find that (a) the resulting motion generates a holomorphic vector field; (b) 
every holomorphic vector field on the state space arises from such a Hamiltonian; (c) 
the symmetry group of the motion is that associated with holomorphically  
projective transformations that map complex geodesics to complex geodesics; and (d) every 
map that preserves complex geodesics on the state space must arise from a Hamiltonian ${\hat K}$ 
that is not necessarily Hermitian. The meaning of these mathematical statements will be explained 
along the way. These results hold in finite as well as in infinite dimensional Hilbert spaces. However, 
for simplicity of exposition we shall be focusing on the case where the Hilbert space is of finite 
(complex) dimension $n$. The associated state space (i.e. the space of rays) is therefore a complex 
projective space ${\mathds C}{\mathbb P}^{n-1}$. Throughout the paper we choose units in which 
$\hbar=1$.
\vspace{0.4cm}

\noindent \textbf{Dynamical equation}. 
Consider a complex Hilbert space ${\mathcal H}$ of complex dimension $n$. A typical element 
of ${\mathcal H}$ is written $|\psiup\rangle$ in the usual Dirac notation. Given a Hamiltonian 
${\hat K}$, Hermitian or not, the dynamical motion on ${\mathcal H}$ is governed by the 
Schr\"odinger equation: 
\begin{eqnarray}
\ri |{\dot\psiup}\rangle = {\hat K} |\psiup\rangle , 
\label{eq:1} 
\end{eqnarray}
where $|{\dot\psiup}\rangle = \partial_t|\psiup\rangle$. The expectation value of an `observable' 
${\hat K}$ (in the generalised sense allowing for complex ones) in the state $|\psiup\rangle$ is 
given by 
\begin{eqnarray}
\langle{\hat K}\rangle = \frac{\langle{\psiup}|{\hat K}|\psiup\rangle}
{\langle{\psiup}|\psiup\rangle} , 
\label{eq:2} 
\end{eqnarray}
where $\langle{\psiup}|$ denotes the Hermitian conjugate of $|\psiup\rangle$. Notice that the 
expectation $\langle{\hat K}\rangle$ is invariant under the complex scale transformation 
$|\psiup\rangle \to \lambda |\psiup\rangle$, $\lambda\in{\mathds C}-\{0\}$. Thus it is sometimes 
convenient to consider the space of an equivalence class of states modulo such transformations. 
This is the space of rays through the origin of ${\mathcal H}$, otherwise known as the complex 
projective space ${\mathds C}{\mathbb P}^{n-1}$. We shall use the term `state space' to mean 
the projective Hilbert space ${\mathds C}{\mathbb P}^{n-1}$. 

The notion of a distance arises naturally from probabilistic considerations of quantum mechanics. 
Specifically, if we write $\rd s$ for the line element on the state space of a neighbouring pair of 
states $|\psiup\rangle$ and $|\psiup+\rd\psiup\rangle=|\psiup\rangle+|\rd\psiup\rangle$, then we 
have the relation
\begin{eqnarray}
\cos^2 {\textstyle\frac{1}{2}}\rd s = \frac{\langle\psiup|\psiup+\rd\psiup\rangle\langle\psiup+
\rd\psiup|\psiup\rangle}
{\langle\psiup|\psiup\rangle\langle\psiup+\rd\psiup|\psiup+\rd\psiup\rangle} 
\label{eq:3}
\end{eqnarray}
for the overlap distance (i.e. the `transition probability' between these neighbouring states). 
Solving this for $\rd s$ and retaining terms of quadratic order, we obtain the following expression 
for the Fubini-Study line element:
\begin{eqnarray}
\rd s^2 = 4 \frac{\langle\psiup|\psiup\rangle\langle\rd\psiup|\rd\psiup\rangle - 
\langle\psiup|\rd\psiup\rangle \langle\rd\psiup|\psiup\rangle}
{\langle\psiup|\psiup\rangle^2} . 
\label{eq:4}
\end{eqnarray}
Wigner's theorem thus can be phrased by saying that the Fubini-Study distance between an 
arbitrary pair of states on the state space is an invariant of motion if and only if the generator 
${\hat K}$ of the dynamics is Hermitian. 

The Schr\"odinger equation (\ref{eq:1}) suffers from the property that $\langle{\psiup}|
{\dot\psiup}\rangle\neq0$, even when ${\hat K}$ is Hermitian; hence it is ill defined on the state 
space. To fix this we shall be considering the modified Schr\"odinger equation: 
\begin{eqnarray}
\ri |{\dot\psiup}\rangle = ({\hat K}-\langle{\hat K}\rangle) |\psiup\rangle , 
\label{eq:5} 
\end{eqnarray}
which is well defined on the state space in the sense that the change in the state is always 
orthogonal to the direction of the state itself. The modified Schr\"odinger equation (\ref{eq:5}) for a 
Hermitian Hamiltonian was introduced by Kibble \cite{kibble} and is often used in the 
analysis of geometric phases, because the solution to (\ref{eq:5}) in ${\mathcal H}$ 
corresponds to the `horizontal lift' of the corresponding trajectory in the state space. 
\vspace{0.4cm}

\noindent \textbf{Geometric quantum mechanics}. 
For further analysis of motions on the state space we are required to employ tools from 
differential geometry. For this purpose, we shall regard the state space, the complex projective 
space ${\mathds C}{\mathbb P}^{n-1}$ of complex dimension $n-1$, as a real manifold 
$\mathfrak{M}$ of real dimension $2n-2$. In this `real' representation (see, for example, Yano \& 
Mogi \cite{yano}) we use roman indices for local tensorial operations in the tangent space of 
$\mathfrak{M}$. A typical point of $\mathfrak{M}$ is written $x^{\boldsymbol a}$, where 
$x^{\boldsymbol a}$ can be thought of as representing a pure state, i.e. an equivalence class 
$\{\lambda |\psiup\rangle\}$, $\lambda\in{\mathds C}-\{0\}$, for some Hilbert space vector 
$|\psiup\rangle$. 
The state space $\mathfrak{M}$ is endowed with the natural line element associated with the 
Fubini-Study metric, which in the real representation will be denoted $\varg_{\boldsymbol{ab}}$. 
Additionally, $\mathfrak{M}$ comes equipped with a natural symplectic structure 
$\omega_{\boldsymbol{ab}}$. The two structures are said to be compatible if there is a complex 
structure $J^{\boldsymbol a}_{\ \boldsymbol b} = \varg^{\boldsymbol{ac}}
\omega_{\boldsymbol{cb}}$, satisfying 
\begin{eqnarray}
J^{\boldsymbol a}_{\ \boldsymbol c} J^{\boldsymbol c}_{\ \boldsymbol b} = 
- \delta^{\boldsymbol a}_{\ \boldsymbol b} 
\label{eq:6}
\end{eqnarray}
such that $\nabla_{\boldsymbol a} J^{\boldsymbol b}_{\ \boldsymbol c} =0$, where 
$\nabla_{\boldsymbol a}$ is the covariant derivative associated with $\varg_{\boldsymbol{ab}}$, and 
$\varg^{\boldsymbol{ac}}\varg_{\boldsymbol{cb}}=\delta^{\boldsymbol a}_{\ \boldsymbol b}$. The 
compatibility condition makes the quantum state space $\mathfrak{M}$ a K\"ahler manifold. Some 
relations that hold among these structures on a K\"ahler manifold useful for calculations that follow 
include: 
$J^{\boldsymbol a}_{\ \boldsymbol c} J^{\boldsymbol b}_{\ \boldsymbol d}\varg_{\boldsymbol{ab}} 
= \varg_{\boldsymbol{cd}}$, 
$J^{\boldsymbol a}_{\ \boldsymbol c} J^{\boldsymbol b}_{\ \boldsymbol d}\omega_{\boldsymbol{ab}} 
= \omega_{\boldsymbol{cd}}$, $\omega_{\boldsymbol{ab}}=\varg_{\boldsymbol{ac}}
J^{\boldsymbol c}_{\ \boldsymbol b}$, $\varg_{\boldsymbol{ab}}=\omega_{\boldsymbol{cb}}
J^{\boldsymbol c}_{\ \boldsymbol a}$, and $\varg_{\boldsymbol{ac}}\omega^{\boldsymbol{bc}}=
J^{\boldsymbol b}_{\ \boldsymbol a}$, where $\omega^{\boldsymbol{ab}} =  \varg^{\boldsymbol{ac}}
\varg^{\boldsymbol{bd}} \omega_{\boldsymbol{cd}}$ is the inverse symplectic structure such that 
we have the relation $\omega_{\boldsymbol{ac}}\omega^{\boldsymbol{bc}}=
\delta_{\boldsymbol a}^{\ \boldsymbol b}$. 

Let us introduce the Hermitian and the skew-Hermitian parts of the Hamiltonian ${\hat K}$ by writing 
${\hat K}={\hat H}-\ri{\hat{\mathit\Gamma}}$, where ${\hat H}^\dagger={\hat H}$ and 
${\hat{\mathit\Gamma}}^\dagger={\hat{\mathit\Gamma}}$ (the minus sign in ${\hat H}-\ri 
{\hat{\mathit\Gamma}}$ is purely conventional). Then the expectation operation (\ref{eq:2}) defines 
a pair of real-valued functions $H(x)$ and ${\mathit\Gamma}(x)$ on $\mathfrak{M}$ according to the 
prescription: 
\begin{eqnarray}
H(x) = \frac{\langle{\psiup}(x)|{\hat H}|\psiup(x)\rangle}{\langle{\psiup}(x)|\psiup(x)\rangle} 
\quad {\rm and} \quad 
{\mathit\Gamma}(x) = \frac{\langle{\psiup}(x)|{\hat{\mathit\Gamma}}|\psiup(x)\rangle}
{\langle{\psiup}(x)|\psiup(x)\rangle}  ,
\label{eq:x2} 
\end{eqnarray}
where $|\psiup(x)\rangle$ represents a state vector in ${\mathcal H}$ associated with the point 
$x^{\boldsymbol a}\in\mathfrak{M}$. The modified Schr\"odingier equation (\ref{eq:5}) is then 
written in the quasi-Hamiltonian form: 
\begin{eqnarray}
\rd x^{\boldsymbol a} = 2\omega^{\boldsymbol{ab}} \nabla_{\boldsymbol b} H(x) \rd t - 
\varg^{\boldsymbol{ab}} \nabla_{\boldsymbol b} {\mathit\Gamma}(x) \rd t,
\label{eq:7}
\end{eqnarray}
which shows that $H$ generates a Hamiltonian symplectic flow, and ${\mathit\Gamma}$ generates 
a Hamiltonian gradient flow. 

The evolution equation (\ref{eq:7}) becomes somewhat trivial if 
$[{\hat H},{\hat{\mathit\Gamma}}]=0$, i.e. if the two operators commute. We note that if $H(x)$ 
and ${\mathit\Gamma}(x)$ are expectations of Hermitian observables ${\hat H}$ and 
${\hat{\mathit\Gamma}}$, then their commutator $\ri[{\hat H},{\hat{\mathit\Gamma}}]$ is also an 
observable, whose expectation is represented on $\mathfrak{M}$ by the Poisson bracket 
$2\omega^{\boldsymbol{ab}} \nabla_{\boldsymbol{a}}H \nabla_{\boldsymbol{b}}{\mathit\Gamma}$.   
Now if $\omega^{\boldsymbol{ab}} \nabla_{\boldsymbol{a}}H 
\nabla_{\boldsymbol{b}}{\mathit\Gamma}\neq0$, then the evolution becomes nontrivial and 
interesting, because of the competition between the Hamiltonian symplectic 
flow and the Hamiltonian gradient flow. In particular, depending on the values of the parameters 
(matrix elements) of the Hamiltonian ${\hat K}$, there can be a `regime change' in the dynamics 
accompanied by a phase transition. In one phase, where $H$ dominates over ${\mathit\Gamma}$ 
in a certain sense, the integral curves of (\ref{eq:7}) possess characteristics of those associated 
with unitary motions, 
whereas in the other phase where ${\mathit\Gamma}$ dominates over $H$, the motions behave 
like normal gradient flows. Such a transition is sometime referred to as a `PT-transition', 
following the work of Bender and Boettcher \cite{BB} on spectral properties of certain complex 
Hamiltonians possessing simultaneous parity and time-reversal symmetry. 
\vspace{0.2cm}

\noindent \textit{Remark}: The symplectic-gradient flow of the form (\ref{eq:7}) on a K\"ahler 
manifold, as arising from the evolution of a quantum state associated with a complex Hamiltonian, 
was first envisaged in Graefe \textit{et al}. \cite{GHK,GS}, and has also been extended 
to mixed-state dynamics in Brody \& Graefe \cite{BG}. For a Hermitian Hamiltonian for which 
${\hat{\mathit\Gamma}}=0$, the fact that the Schr\"odingier equation can be expressed as  
Hamilton's equations of classical mechanics was first observed by Dirac \cite{dirac}, and 
subsequently by Strocchi \cite{strocchi}. This observation led Kibble \cite{kibble} and Weinberg 
\cite{weinberg} to generalise quantum mechanics into the nonlinear domain, where the Hamiltonian 
function $H(x)$ on the state space does not take the special form (\ref{eq:x2}). Formulation of 
standard quantum theory in the language of projective geometry was initiated by Kibble 
\cite{kibble}, but also to some extent by Cantoni \cite{cantoni,cantoni2}, and investigated further 
by many authors, including, in particular, Heslot \cite{heslot}, Page \cite{page}, 
Cirelli \textit{et al}. \cite{italian}, Anandan \& Aharanov \cite{AA}, Gibbons \cite{GWG}, Ashtekar 
\& Schilling \cite{AS}, Hughston \cite{LPH,LPH2}, to name a few. For a more comprehensive list of 
references, see Brody \& Hughston \cite{BH} and Bengtsson \& \.Zyczkowski \cite{polish}. 
\vspace{0.2cm}

\noindent \textit{Remark}: The factor of two appearing in (\ref{eq:7}) is due to the fact that the 
`quantum symplectic structure' is one-half of the natural symplectic structure on ${\mathfrak M}$. 
As explained in Hughston \cite{LPH}, the natural symplectic structure is selected on account of 
the condition that $\omega^{\boldsymbol{ab}} =  \varg^{\boldsymbol{ac}}
\varg^{\boldsymbol{bd}} \omega_{\boldsymbol{cd}}$ gives the inverse symplectic structure. The 
quantum symplectic structure appearing in the Schr\"odingier equation, as well as in the Poisson 
bracket, however, is given by $\frac{1}{2}\omega_{\boldsymbol{ab}}$, whose inverse is thus 
$2\omega^{\boldsymbol{ab}}$. 
\vspace{0.2cm}

When ${\hat{\mathit\Gamma}}=0$ so that ${\hat K}={\hat H}$ is Hermitian, the evolution equation 
(\ref{eq:7}) generates a Hamiltonian vector field $\xi^{\boldsymbol{a}}=2\omega^{\boldsymbol{ab}} 
\nabla_{\boldsymbol b} H$ which satisfies the Killing equation 
\begin{eqnarray}
\nabla_{({\boldsymbol a}}\xi_{{\boldsymbol b})} = 0 , 
\label{eq:8} 
\end{eqnarray}
where $\xi_{\boldsymbol a} = \varg_{\boldsymbol{ab}}\xi^{\boldsymbol b}$. 
An alternative way of expressing 
this is to write ${\mathfrak L}_\xi \varg_{\boldsymbol{ab}}=0$, that is, the Lie derivative (cf. Yano 
\cite{yano3}) of the Fubini-Study metric associated with the vector filed $\xi^{\boldsymbol{a}}$ 
vanishes. Conversely, given a Killing field 
$\xi^{\boldsymbol{a}}$, the associated quantum (Hermitian) Hamiltonian can be recovered 
according to the relation  
\begin{eqnarray}
\half \omega^{\boldsymbol{ab}} \nabla_{\boldsymbol{a}}\xi_{\boldsymbol{b}} = n (H-{\bar H}),  
\label{eq:9} 
\end{eqnarray}
where ${\bar H}={\rm tr}({\hat H})/n$ is the uniform average of the eigenvalues of ${\hat H}$. To 
see this, we remark first that if we define as usual the Laplace-Beltrami operator $\nabla^2$ on 
$\mathfrak{M}$ by $\nabla^2 = \varg^{\boldsymbol{ab}}\nabla_{\boldsymbol{a}}
\nabla_{\boldsymbol{b}}$, then we have 
\begin{eqnarray}
\nabla^2 H = n({\bar H}-H).   
\label{eq:10} 
\end{eqnarray}
This can easily be verified if we lift the calculation to the Hilbert space ${\mathcal H}$, perform 
the calculation using the form (\ref{eq:x2}) of a quantum observable, and unit normalise the states 
after the calculation. We thus find 
\begin{eqnarray}
\half \omega^{\boldsymbol{ab}} \nabla_{\boldsymbol{a}}\xi_{\boldsymbol{b}} &=& 
\omega^{\boldsymbol{ab}} \nabla_{\boldsymbol{a}} (\varg_{\boldsymbol{bd}} 
\omega^{\boldsymbol{dc}} \nabla_{\boldsymbol{c}}H) \nonumber \\ &=& -  
\omega^{\boldsymbol{ab}} \omega^{\boldsymbol{cd}} \varg_{\boldsymbol{bd}}
\nabla_{\boldsymbol{a}} \nabla_{\boldsymbol{c}}H \nonumber \\ &=& -\nabla^2 H ,
\label{eq:11} 
\end{eqnarray}
and hence (\ref{eq:9}). This is essentially the geometric derivation of Wigner's theorem, showing 
that a unitary motion generates an isometry of the state space, and that an isometry of the state 
space must arise from a generator that is Hermitian. 
\vspace{0.2cm}

\noindent \textit{Remark}: Viewed as a mathematical statement, a form of Wigner's theorem was 
in fact observed earlier by Mannoury \cite{mannoury} in the context of embedding complex 
projective spaces in Euclidean spaces. The analysis of Mannoury was extended further by Hodge 
\cite{hodge}, with the following construction. Let $|\psiup\rangle = (\psi^1,\psi^2,\cdots,\psi^n)$ be 
the homogeneous coordinates for ${\mathds C}{\mathbb P}^{n-1}$, with the convention 
$\langle\psiup|\psiup\rangle=1$, and introduce coordinates on ${\mathds R}^{n^2}$ by 
$(x^k, x^{hk}, y^{hk})$, where $h,k=1,\ldots,n$, $k\neq h$. We can isometrically embed 
${\mathds C}{\mathbb P}^{n-1}$ in ${\mathds R}^{n^2}$ according to the prescription: 
\begin{eqnarray}
x^h = \sqrt{2}\psi^h{\bar\psi}^h, \quad x^{hk}=\psi^h{\bar\psi}^k+\psi^k{\bar\psi}^h, \quad 
y^{hk}=\ri(\psi^h{\bar\psi}^k-\psi^k{\bar\psi}^h) . 
\label{eq:12} 
\end{eqnarray}
The image then lies on a sphere $S^{n^2-2}$ in a hyperplane ${\mathds R}^{n^2-1}$ defined by 
the linear equation $x^1+x^2+\cdots+x^n=\sqrt{2}$. In particular, for $n=2$ the result is just the 
`Bloch sphere' of two-level systems in quantum mechanics. For $n>2$, this technique might prove 
useful for analysing the space of mixed states in quantum mechanics. As Kobayashi \cite{kobayashi} 
points out, the construction (\ref{eq:12}) gives a minimal embedding of 
${\mathds C}{\mathbb P}^{n-1}$ in a Euclidean space. The metric geometry of 
${\mathds C}{\mathbb P}^{n-1}$ induced by the ambient Euclidean geometry can then be worked 
out (see Hodge \cite{hodge} for detail), which provides another way of deriving the Fubini-Study 
geometry. A natural question thus arising concerns transformations that leave the metric invariant; 
in this context, Mannoury \cite{mannoury} found that this is given by `conjugate orthogonal' 
(i.e. unitary) transformations. 
\vspace{0.2cm}

Let us introduce some identities in relation to motions on ${\mathfrak M}$ generated by Hermitian 
Hamiltonians that will be useful in the ensuing analysis. If $\xi^{\boldsymbol{a}}$ satisfies the 
Killing equation $\nabla_{({\boldsymbol b}}\xi_{{\boldsymbol c})} = 0$, then upon differentiation 
we have 
$\nabla_{{\boldsymbol a}}\nabla_{{\boldsymbol b}}\xi_{{\boldsymbol c}}+
\nabla_{{\boldsymbol a}}\nabla_{{\boldsymbol c}}\xi_{{\boldsymbol b}}=0$, and hence, by 
interchanging the index pair $({\boldsymbol a},{\boldsymbol c})$ and taking the difference, we 
obtain $\nabla_{[{\boldsymbol a}}\nabla_{|{\boldsymbol b}|}\xi_{{\boldsymbol c}]}=0$. On account 
of the Ricci identity 
\begin{eqnarray}
\nabla_{\boldsymbol b}\nabla_{\boldsymbol a}\xi_{\boldsymbol c}-
\nabla_{\boldsymbol a}\nabla_{\boldsymbol b}\xi_{\boldsymbol c} = 
R_{\boldsymbol{abc}}^{\ \ \ \ {\boldsymbol d}}\xi_{\boldsymbol d}, 
\label{eq:z14}
\end{eqnarray}
which holds for any vector field $\xi^{\boldsymbol{a}}$, and the cyclic identity 
\begin{eqnarray}
R_{\boldsymbol{abc}}^{\ \ \ \ {\boldsymbol d}} + R_{\boldsymbol{bca}}^{\ \ \ \ {\boldsymbol d}}+
R_{\boldsymbol{cab}}^{\ \ \ \ {\boldsymbol d}} = 0
\label{eq:zz14}
\end{eqnarray}
for the Riemann tensor, we thus obtain 
\begin{eqnarray}
\nabla_{\boldsymbol c}\nabla_{\boldsymbol a}\xi_{\boldsymbol b} = 
R_{\boldsymbol{abc}}^{\ \ \ \ {\boldsymbol d}}\xi_{\boldsymbol d} , 
\label{eq:x14}
\end{eqnarray}
a standard identity in Riemannian geometry for a Killing field. If we substitute the expression 
$\xi_{\boldsymbol a} = 2J_{\boldsymbol a}^{\ \boldsymbol b}\nabla_{\boldsymbol b}H$ in 
(\ref{eq:x14}) we find 
\begin{eqnarray}
\nabla_{\boldsymbol c}\nabla_{\boldsymbol a}\nabla_{\boldsymbol b}H = - 
R_{\boldsymbol{apc}}^{\ \ \ \ {\boldsymbol q}} 
J_{\ {\boldsymbol b}}^{\boldsymbol p} J_{\ {\boldsymbol q}}^{\boldsymbol d} 
\nabla_{\boldsymbol d}H .
\end{eqnarray}
This identity, which has been obtained in the context of quantum observables, for example, in 
Cirelli \textit{et al}. \cite{italian} and in Hughston \cite{LPH2}, holds 
when $H(x)$ is the expectation of a Hermitian Hamiltonian ${\hat H}$. It follows from the 
expression 
\begin{eqnarray}
R_{\boldsymbol{apc}}^{\ \ \ \ {\boldsymbol q}} = - \quat \Big( \varg_{\boldsymbol{ac}} 
\delta_{{\boldsymbol p}}^{\ \boldsymbol q} - \varg_{\boldsymbol{pc}} 
\delta_{{\boldsymbol a}}^{\ \boldsymbol q} - \omega_{\boldsymbol{ac}} 
J_{{\boldsymbol p}}^{\ \boldsymbol q} + \omega_{\boldsymbol{pc}} 
J_{{\boldsymbol a}}^{\ \boldsymbol q} - 2 \omega_{\boldsymbol{ap}} 
J_{{\boldsymbol c}}^{\ \boldsymbol q} \Big)
\label{eq:x16} 
\end{eqnarray}
for the Riemann tensor on a Fubini-Study manifold that 
\begin{eqnarray} 
\nabla_{\boldsymbol c}\nabla_{\boldsymbol a}\nabla_{\boldsymbol b}H &=& -\quat \Big( 
2 \varg_{\boldsymbol{ab}} \nabla_{\boldsymbol c}H +
\varg_{\boldsymbol{bc}} \nabla_{\boldsymbol a}H +  
\varg_{\boldsymbol{ca}} \nabla_{\boldsymbol b}H  \nonumber \\ && \qquad \qquad \qquad 
+\omega_{\boldsymbol{cb}} J_{{\boldsymbol a}}^{\ \boldsymbol d} \nabla_{\boldsymbol d}H + 
\omega_{\boldsymbol{ca}} J_{{\boldsymbol b}}^{\ \boldsymbol d} \nabla_{\boldsymbol d}H 
\Big) .
\label{eq:x17} 
\end{eqnarray} 
Contracting the index pair $({\boldsymbol a},{\boldsymbol b})$, we thus obtain 
\begin{eqnarray}
\nabla_{\boldsymbol c}\nabla^2 H =-n \nabla_{\boldsymbol c}H, 
\end{eqnarray}
which provides another 
derivation for (\ref{eq:10}). Relation (\ref{eq:x17}) appears in Yano \& Hiramatu 
\cite{yano2} in the context of K\"ahler manifolds with constant positive curvatures. Specifically, 
if the system of partial differential equations (\ref{eq:x17}) admits a nontrivial solution $H(x)$, 
then the manifold is necessarily a complex projective space equipped with a Fubini-Study metric 
of constant holomorphic sectional curvature. The identity (\ref{eq:x17}) for quantum observables 
plays an important role in what follows in establishing the main result of this paper. 
\vspace{0.4cm}

\noindent \textbf{Projective transformations}. 
We saw how, from a geometric point of view, Wigner's theorem can be interpreted as stating 
that the motion generated by a Hermitian Hamiltonian ${\hat H}$ gives rise to Killing fields on the 
state space. As indicated above, the purpose of the present paper is to show that when ${\hat H}$ 
is replaced by a complex Hamiltonian ${\hat K}$, the resulting dynamics generate holomorphically  
projective fields. The notion of a projective transformation on a Riemannian manifold is perhaps not 
so widely appreciated in physics literature, so it will be useful to briefly introduced the idea here. 
The concept of a \textit{holomorphically} projective transformation should then become more 
transparent. 

There are various transformations on a Riemannian manifold, such as Killing motions or 
conformal motions, that are of interests in a variety of contexts. A projective transformation 
corresponds to the most general motion that maps geodesics to geodesics. 
Recall that a geodesic curve $x^{\boldsymbol a}(s)$ on a Riemannian (or K\"ahlerian) manifold 
${\mathfrak M}$ is characterised by the fact that when a tangent vector $u^{\boldsymbol a} = 
\rd x^{\boldsymbol a}/\rd s$ is parallel transported along the curve, then it remains tangent to the 
curve. In other words, we have $\nabla_u u \propto u$, i.e. $u^{\boldsymbol a} 
\nabla_{\boldsymbol a} u^{\boldsymbol b} = \alpha(s) u^{\boldsymbol b}$ for some real scalar 
function $\alpha(s)$. Multiplying both sides by $u^{\boldsymbol c}$, we see that the right side is 
symmetric in the index pair $({\boldsymbol b},{\boldsymbol c})$, which means that the 
antisymmetric part of the left side equals zero: 
\begin{eqnarray}
(u^{\boldsymbol a} \nabla_{\boldsymbol a} u^{[{\boldsymbol b}}) u^{{\boldsymbol c}]}=0.
\end{eqnarray}
Consider now the effect of dragging a geodesic curve along the integral curve of a vector field 
$\xi^{\boldsymbol a}$. If ${\hat\nabla}_{\boldsymbol a}$ is the resulting transported Levi-Civita 
connection, and if geodesic curves remain geodesic curves under the transportation, then we 
must have $(u^{\boldsymbol a} {\hat\nabla}_{\boldsymbol a} u^{[{\boldsymbol b}}) 
u^{{\boldsymbol c}]}=0$, or equivalently, $u^{\boldsymbol a} u^{\boldsymbol b}({\mathfrak L}_\xi 
\Gamma_{\boldsymbol{ab}}^{[{\boldsymbol c}}) u^{{\boldsymbol d}]}=0$, where 
$\Gamma_{\boldsymbol{ab}}^{{\boldsymbol c}}$ denotes the Christoffel symbol. This condition is 
satisfied if and only if (see, e.g., Yano \cite{yano3}) there exists a vector field $\phi_{\boldsymbol a}$ such that ${\mathfrak L}_\xi \Gamma_{\boldsymbol{ab}}^{{\boldsymbol c}} = 
\phi_{({\boldsymbol a}} \delta_{\boldsymbol{b})}^{\ \boldsymbol c}$, that is, 
\begin{eqnarray}
\nabla_{\boldsymbol a}\nabla_{\boldsymbol b}\xi^{\boldsymbol c} + 
R^{\boldsymbol c}_{\ \boldsymbol{bad}}\xi^{\boldsymbol d}  = 
\phi_{({\boldsymbol a}} \delta_{\boldsymbol{b})}^{\ \boldsymbol c} \ , 
\label{eq:x19} 
\end{eqnarray}
where we have made use of the symmetry properties 
$R_{\boldsymbol{abcd}}=R_{[\boldsymbol{ab}][\boldsymbol{cd}]}$ and 
$R_{\boldsymbol{abcd}}=R_{\boldsymbol{cdab}}$ of the Riemann tensor. 

Relation (\ref{eq:x19}) gives the necessary and sufficient condition that the vector field 
$\xi^{\boldsymbol a}$ preserves geodesics. It should be evident that for a Killing field, for which 
${\mathfrak L}_\xi \Gamma_{\boldsymbol{ab}}^{{\boldsymbol c}} = 0$ holds, (\ref{eq:x19}) is 
automatically satisfied with $\phi_{\boldsymbol a}=0$. More generally, if we contract the index pair 
$({\boldsymbol b},{\boldsymbol c})$ in (\ref{eq:x19}), then we obtain 
\begin{eqnarray}
\phi_{\boldsymbol a} = \frac{1}{2n-1}\, 
\nabla_{\boldsymbol a}\nabla_{\boldsymbol b}\xi^{\boldsymbol b} , 
\label{eq:x20}
\end{eqnarray}
which shows that $\phi_{\boldsymbol a}$ is a necessarily a gradient vector. 
\vspace{0.4cm}

\noindent \textbf{Complex geodesics}. 
On account of the importance of geodesic curves in various applications, geodesics-preserving 
maps have been investigated extensively in the literature. For a K\"ahler manifold, however, 
conditions (\ref{eq:x19}) turn out to be somewhat overly stringent, and often provide no nontrivial 
solution other than Killing (for which ${\mathfrak L}_\xi \varg_{\boldsymbol{ab}}=0$) or affine (for 
which ${\mathfrak L}_\xi \Gamma_{\boldsymbol{bc}}^{{\boldsymbol a}} =0$) transformations. This 
motivated Otsuki \& Tashiro \cite{otsuki} and Tashiro \cite{tashiro} to introduce the notion of 
`holomorphically planer curves', otherwise known as complex geodesics. 

Recall the condition $\nabla_u u \propto u$ for a geodesic curve that a tangent vector transported 
parallelly along the curve remains tangent to the curve. Suppose that we relax this condition slightly 
by demanding that a special tangent two-plane is parallel transported along the curve into a tangent 
two-plane of the same type. Specifically, if $u^{\boldsymbol a} = \rd x^{\boldsymbol a}/\rd s$ is a 
tangent vector of the curve $x^{\boldsymbol a}(s)$ at $s$, then we can use the complex structure 
to form another vector $J^{{\boldsymbol a}}_{\ \boldsymbol b} u^{\boldsymbol b}$ orthogonal to 
$u^{\boldsymbol a}$. The pair of vectors 
$(u^{\boldsymbol a},J^{{\boldsymbol a}}_{\ \boldsymbol b} u^{\boldsymbol b})$ then span a 
holomorphic two-plane (section) tangent to the curve. If this two-plane is parallel transported along 
the curve in such a manner that the plane remains a holomorphic tangent plane, then we must have 
\begin{eqnarray} 
u^{\boldsymbol a} \nabla_{\boldsymbol a} u^{\boldsymbol b} = \left( \alpha(s) \, 
\delta^{\boldsymbol b}_{\ \boldsymbol c}  + \beta(s)\, J^{{\boldsymbol b}}_{\ \boldsymbol c} \right) 
u^{\boldsymbol c}
\label{eq:x21} 
\end{eqnarray} 
for a pair of real functions $\alpha(s)$ and $\beta(s)$. Equivalently, in terms of 
$x^{\boldsymbol a}(s)$ we have 
\begin{eqnarray} 
\frac{\rd^2 x^{\boldsymbol a}}{\rd s^2} + \Gamma_{\boldsymbol{bc}}^{{\boldsymbol a}} 
\frac{\rd x^{\boldsymbol b}}{\rd s} \frac{\rd x^{\boldsymbol c}}{\rd s} = \left( \alpha(s) \, 
\delta^{\boldsymbol a}_{\ \boldsymbol b}  + \beta(s)\, J^{{\boldsymbol a}}_{\ \boldsymbol b} \right) 
\frac{\rd x^{\boldsymbol b}}{\rd s} ,
\label{eq:x22} 
\end{eqnarray} 
which is the defining equation for a holomorphically planer curve. Since the complex structure 
is a real representation for the multiplication by a unit imaginary number, we can think of 
(\ref{eq:x22}) as a `geodesic' equation for which the real proportionality factor $\alpha(s)$ is 
replaced by a complex factor $\alpha(s)+\ri\beta(s)$. It is for this reason that (\ref{eq:x22}) is 
sometime informally referred to as a complex geodesic equation (cf. Otsuki \& Tashiro 
\cite{otsuki}, \S8). 

Suppose that we drag a holomorphically planer curve along a vector field $\xi^{\boldsymbol a}$. 
If the resulting curve remains holomorphically planer, then writing 
${\hat\Gamma}_{\boldsymbol{bc}}^{{\boldsymbol a}}$ for the dragged Christoffel symbol, we must 
have 
\begin{eqnarray} 
\frac{\rd^2 x^{\boldsymbol a}}{\rd s^2} + {\hat\Gamma}_{\boldsymbol{bc}}^{{\boldsymbol a}} 
\frac{\rd x^{\boldsymbol b}}{\rd s} \frac{\rd x^{\boldsymbol c}}{\rd s} = \left( \alpha'(s) \, 
\delta^{\boldsymbol a}_{\ \boldsymbol b}  + \beta'(s)\, J^{{\boldsymbol a}}_{\ \boldsymbol b} \right) 
\frac{\rd x^{\boldsymbol b}}{\rd s} 
\label{eq:x23} 
\end{eqnarray} 
for some real functions $\alpha'$ and $\beta'$. This condition is fulfilled if and only if there exists 
a vector field $\phi_{\boldsymbol a}$ such that ${\hat\Gamma}_{\boldsymbol{bc}}^{{\boldsymbol a}} 
- \Gamma_{\boldsymbol{bc}}^{{\boldsymbol a}}={\mathfrak L}_\xi 
\Gamma_{\boldsymbol{ab}}^{{\boldsymbol c}}$ can be expressed in the form:
\begin{eqnarray}
{\mathfrak L}_\xi \Gamma_{\boldsymbol{ab}}^{{\boldsymbol c}} = 
\nabla_{\boldsymbol a}\nabla_{\boldsymbol b}\xi^{\boldsymbol c} + 
\xi^{\boldsymbol d} R_{\boldsymbol{dba}}^{\ \ \ \ \boldsymbol c}  = 
\phi_{{\boldsymbol a}} \delta_{\boldsymbol{b}}^{\ \boldsymbol c} + 
\phi_{{\boldsymbol b}} \delta_{\boldsymbol{a}}^{\ \boldsymbol c} - 
\phi_{{\boldsymbol d}} J^{{\boldsymbol d}}_{\ \boldsymbol b}J^{{\boldsymbol c}}_{\ \boldsymbol a}
-\phi_{{\boldsymbol d}}J^{{\boldsymbol d}}_{\ \boldsymbol a}J^{{\boldsymbol c}}_{\ \boldsymbol b}.
\label{eq:x24} 
\end{eqnarray}
Contracting the index pair $({\boldsymbol b},{\boldsymbol c})$, and using 
$R_{\boldsymbol{dba}}^{\ \ \ \ \boldsymbol b}=0$ and $J^{{\boldsymbol b}}_{\ \boldsymbol b}=0$, 
we find 
\begin{eqnarray}
\phi_{\boldsymbol a} = \frac{1}{2n}\, 
\nabla_{\boldsymbol a}\nabla_{\boldsymbol b}\xi^{\boldsymbol b} , 
\label{eq:x25}
\end{eqnarray}
which shows that $\phi_{\boldsymbol a}$ is necessarily a gradient vector. On the other hand, 
transacting (\ref{eq:x25}) with $\varg^{\boldsymbol{ab}}$, we obtain 
\begin{eqnarray}
\nabla^2 \nabla_{\boldsymbol c} + \xi^{\boldsymbol d} R_{\boldsymbol{d}}^{\ \ \boldsymbol c}=0,
\label{eq:x26}
\end{eqnarray}
which shows that $\xi^{\boldsymbol a}$ is necessarily a holomorphic vector satisfying 
${\mathfrak L}_\xi J^{{\boldsymbol c}}_{\ \boldsymbol b}=0$, that is, 
$J^{{\boldsymbol c}}_{\ \boldsymbol a} \nabla_{\boldsymbol b} \xi_{\boldsymbol c} + 
J^{{\boldsymbol c}}_{\ \boldsymbol b} \nabla_{\boldsymbol c} \xi_{\boldsymbol a}=0$, or 
equivalently, 
\begin{eqnarray}
(\nabla_{\boldsymbol d} \xi_{\boldsymbol c}) J^{{\boldsymbol d}}_{\ \boldsymbol b}
J^{{\boldsymbol c}}_{\ \boldsymbol a} = \nabla_{\boldsymbol b} \xi_{\boldsymbol c} .  
\label{eq:x27}
\end{eqnarray}
This follows on account of the fact that (\ref{eq:x26}) is a necessary and sufficient condition that 
the analyticity condition (\ref{eq:x27}) holds (see Yano \cite{yano3}). 
\vspace{0.2cm}

\noindent \textit{Remark}: Following on the work of Otsuki \& Tashiro \cite{otsuki} and Tashiro 
\cite{tashiro}, properties of holomorphically projective transformations characterised by the 
condition (\ref{eq:x24}) on K\"ahler manifolds have been investigated by various authors, including, 
in particular, Tachibana \& Ishihara \cite{ishihara,ishihara2}, Yoshimatsu, \cite{yoshimatsu}, and 
Yano \& Hiramatu \cite{yano2}. See Yano \cite{yano4}, \S XII, for a textbook exposition of the subject. 
More recently, the subject has gained renewed interests in relation to the notion of Hamiltonian 
two-forms of Apostolov \textit{et al}. \cite{apostolov}; see Matveev \& Rosemann \cite{matveev} 
for further details on this connection. 
\vspace{0.4cm}

\noindent \textbf{Complex dynamics}. 
Returning to the complexified Schr\"odinger dynamics (\ref{eq:7}), let us now establish that the 
associated vector field 
\begin{eqnarray}
\xi^{\boldsymbol a} = 2\omega^{\boldsymbol{ab}} \nabla_{\boldsymbol b} H - 
\varg^{\boldsymbol{ab}} \nabla_{\boldsymbol b} {\mathit\Gamma}
\label{eq:x28}
\end{eqnarray}
fulfils the condition (\ref{eq:x24}) for preserving complex geodesics on the quantum state space 
${\mathfrak M}$. Since the Killing field $2\omega^{\boldsymbol{ab}} \nabla_{\boldsymbol b} H$ 
satisfies ${\mathfrak L}_\xi \Gamma_{\boldsymbol{bc}}^{{\boldsymbol a}} =0$ on account of 
(\ref{eq:x14}), it suffices to focus attention on the term $-\varg^{\boldsymbol{ab}} 
\nabla_{\boldsymbol b} {\mathit\Gamma}$. Then a calculation shows that 
\begin{eqnarray}
\nabla_{\boldsymbol a}\nabla_{\boldsymbol b}\xi_{\boldsymbol c} + 
R_{\boldsymbol{cbad}} \xi^{\boldsymbol d} &=& -\nabla_{\boldsymbol a}\nabla_{\boldsymbol b}
\nabla_{\boldsymbol c} {\mathit\Gamma} - R_{\boldsymbol{cbad}} \nabla^{\boldsymbol d} 
{\mathit\Gamma}  \nonumber \\ &=& -R_{\boldsymbol{apbq}} J^{{\boldsymbol p}}_{\ \boldsymbol r} 
\nabla^{\boldsymbol r} {\mathit\Gamma} J^{{\boldsymbol q}}_{\ \boldsymbol c} - 
R_{\boldsymbol{cbad}} \nabla^{\boldsymbol d} {\mathit\Gamma} \nonumber \\ &=& 2 \left( 
\varg_{\boldsymbol{ca}} \nabla_{\boldsymbol b} {\mathit\Gamma} + 
\varg_{\boldsymbol{cb}} \nabla_{\boldsymbol a} {\mathit\Gamma} - 
\omega_{\boldsymbol{ca}} J^{{\boldsymbol d}}_{\ \boldsymbol b}
\nabla_{\boldsymbol d} {\mathit\Gamma} -
\omega_{\boldsymbol{cb}} J^{{\boldsymbol d}}_{\ \boldsymbol a}
\nabla_{\boldsymbol d} {\mathit\Gamma}  \right) , 
\end{eqnarray}
where we have made use of the expression (\ref{eq:x16}) for the Riemann tensor on 
${\mathfrak M}$. This relation agrees with (\ref{eq:x24}), with $\phi_{\boldsymbol a} = 2 \nabla_{\boldsymbol a} {\mathit\Gamma}$, thus establishing the claim (c) that evolution equation on 
${\mathfrak M}$ generated by a complex Hamiltonian ${\hat K}={\hat H}-\ri{\hat{\mathit\Gamma}}$ 
gives rise to a holomorphically projective transformation. 

To proceed, let us follow closely the argument of Tachibana \& Ishihara \cite{ishihara} and 
derive an integrability condition for (\ref{eq:x24}) by examining the deviation of 
the Riemann tensor. For this purpose we make use of the following identity (see Yano \cite{yano4}): 
\begin{eqnarray}
{\mathfrak L}_\xi R_{\boldsymbol{dba}}^{\ \ \ \ \boldsymbol c} = \nabla_{\boldsymbol d} 
{\mathfrak L}_\xi \Gamma_{\boldsymbol{ba}}^{{\boldsymbol c}} - \nabla_{\boldsymbol b} 
{\mathfrak L}_\xi \Gamma_{\boldsymbol{da}}^{{\boldsymbol c}} . 
\label{eq:x30}
\end{eqnarray}
Substituting (\ref{eq:x24}) in (\ref{eq:x30}) and rearranging terms we obtain 
\begin{eqnarray}
{\mathfrak L}_\xi R_{\boldsymbol{dba}}^{\ \ \ \ \boldsymbol c} &=& 
\delta^{{\boldsymbol c}}_{\ \boldsymbol b}\nabla_{\boldsymbol d} \phi_{\boldsymbol a} - 
\delta^{{\boldsymbol c}}_{\ \boldsymbol d}\nabla_{\boldsymbol b} \phi_{\boldsymbol a} 
\nonumber \\ && - J^{{\boldsymbol c}}_{\ \boldsymbol b}
\nabla_{\boldsymbol d}J^{{\boldsymbol p}}_{\ \boldsymbol a}
\phi_{\boldsymbol p} + J^{{\boldsymbol c}}_{\ \boldsymbol d}\nabla_{\boldsymbol b}
J^{{\boldsymbol p}}_{\ \boldsymbol a}\phi_{\boldsymbol p} - \left( 
\nabla_{\boldsymbol d} J^{{\boldsymbol p}}_{\ \boldsymbol b}\phi_{\boldsymbol p} - 
\nabla_{\boldsymbol b} J^{{\boldsymbol p}}_{\ \boldsymbol d}\phi_{\boldsymbol p} \right) 
J^{{\boldsymbol c}}_{\ \boldsymbol a}, 
\label{eq:x31}
\end{eqnarray}
where we have made use of the fact that $\phi_{\boldsymbol a}$ is a gradient vector, satisfying 
$\nabla_{[\boldsymbol{b}} \phi_{\boldsymbol{d}]}=0$. The relation (\ref{eq:x31}) is the desired 
integrability condition. In particular, contracting the index pair $({\boldsymbol c},{\boldsymbol d})$, 
we thus deduce that 
\begin{eqnarray}
{\mathfrak L}_\xi R_{\boldsymbol{ba}} = -(2n-2) \nabla_{\boldsymbol b} \phi_{\boldsymbol a} 
- 2 J^{{\boldsymbol p}}_{\ \boldsymbol b} J^{{\boldsymbol q}}_{\ \boldsymbol a} 
\nabla_{\boldsymbol p} \phi_{\boldsymbol q} , 
\label{eq:x32}
\end{eqnarray}
where we have made use of the fact that $\nabla_{\boldsymbol b} \phi_{\boldsymbol a} = 
\nabla_{\boldsymbol{a}}\phi_{\boldsymbol{b}}$, i.e. $\phi_{\boldsymbol a}$ is a gradient vector. 
On the other hand, from the fact that $\xi^{\boldsymbol a}$ is analytic, we have 
\begin{eqnarray}
{\mathfrak L}_\xi R_{\boldsymbol{ba}} = ({\mathfrak L}_\xi R_{\boldsymbol{pq}} ) 
J^{{\boldsymbol p}}_{\ \boldsymbol b} J^{{\boldsymbol q}}_{\ \boldsymbol a} .
\label{eq:x33}
\end{eqnarray}
Putting together (\ref{eq:x32}) and (\ref{eq:x33}) we thus find 
\begin{eqnarray}
\nabla_{\boldsymbol b} \phi_{\boldsymbol a}  = J^{{\boldsymbol p}}_{\ \boldsymbol b} 
J^{{\boldsymbol q}}_{\ \boldsymbol a} \nabla_{\boldsymbol p} \phi_{\boldsymbol q} , 
\label{eq:x34}
\end{eqnarray}
which on account of (\ref{eq:x27}) shows that $\phi_{\boldsymbol a}$ is analytic. Furthermore, 
it also follows from (\ref{eq:x34}) that 
\begin{eqnarray}
\nabla_{\boldsymbol b}(J^{{\boldsymbol c}}_{\ \boldsymbol a}\phi_{\boldsymbol c}) + 
\nabla_{\boldsymbol a}(J^{{\boldsymbol c}}_{\ \boldsymbol b}\phi_{\boldsymbol c}) &=& 
J^{{\boldsymbol c}}_{\ \boldsymbol a}\nabla_{\boldsymbol b}\phi_{\boldsymbol c} + 
J^{{\boldsymbol c}}_{\ \boldsymbol b} J^{{\boldsymbol p}}_{\ \boldsymbol a} 
J^{{\boldsymbol q}}_{\ \boldsymbol c} \nabla_{\boldsymbol p}\phi_{\boldsymbol q} 
\nonumber \\ &=& J^{{\boldsymbol c}}_{\ \boldsymbol a}\nabla_{\boldsymbol b}
\phi_{\boldsymbol c} - J^{{\boldsymbol c}}_{\ \boldsymbol a}\nabla_{\boldsymbol c}
\phi_{\boldsymbol b} \nonumber \\ &=& 0, 
\label{eq:x35}
\end{eqnarray}
since $\nabla_{\boldsymbol b} \phi_{\boldsymbol a} = 
\nabla_{\boldsymbol{a}}\phi_{\boldsymbol{b}}$. In other words, 
$J^{{\boldsymbol c}}_{\ \boldsymbol a}\phi_{\boldsymbol c}$ is a Killing vector. 

From the geometric characterisation of Wigner's theorem, however, a Killing field 
$J^{{\boldsymbol c}}_{\ \boldsymbol a}\phi_{\boldsymbol c}$ on the quantum state space 
${\mathfrak M}$ is necessarily generated by a quantum observable ${\mathit\Gamma}(x)$ of the 
form (\ref{eq:x2}) such that 
\begin{eqnarray}
\phi_{\boldsymbol a} = - \half 
\nabla_{\boldsymbol a} {\mathit\Gamma} ,
\label{eq:x36}
\end{eqnarray}
where the factor of $-\frac{1}{2}$ is purely conventional. Since, up to an additive constant, such a 
${\mathit\Gamma}$ is an eigenfunction of the Laplacian, i.e. $\nabla_{\boldsymbol a} \nabla^2 
{\mathit\Gamma} = -n \nabla_{\boldsymbol a} {\mathit\Gamma}$, we find that 
\begin{eqnarray}
\phi_{\boldsymbol a} = - \frac{1}{2n}\, 
\nabla_{\boldsymbol a} \nabla_{\boldsymbol b} \nabla^{\boldsymbol b} {\mathit\Gamma} . 
\label{eq:x37}
\end{eqnarray}
Comparing (\ref{eq:x37}) and (\ref{eq:x25}) we thus deduce that $\xi^{\boldsymbol a}$ must be 
expressible in the from (\ref{eq:x28}), where $H(x)$ and ${\mathit\Gamma}(x)$ are necessarily 
of the form (\ref{eq:x2}). In particular, given a holomorphically projective transformation 
$\xi^{\boldsymbol a}$ on the quantum state space, the corresponding Hermitian and skew-Hermitian 
parts of the Hamiltonian can be recovered according to the prescription: 
\begin{eqnarray}
H-{\bar H} = \frac{1}{2n}\, \omega^{\boldsymbol{ab}} \nabla_{\boldsymbol{a}}\xi_{\boldsymbol{b}} 
\quad {\rm and} \quad 
{\mathit\Gamma}-{\bar{\mathit\Gamma}} = -\frac{1}{n}\, \varg^{\boldsymbol{ab}} 
\nabla_{\boldsymbol{a}}\xi_{\boldsymbol{b}} . 
\label{eq:x38} 
\end{eqnarray}
This completes the verification of the claim (d) that a holomorphically projective transformation 
necessarily arises from a Hamiltonian ${\hat K}$ that is not necessarily Hermitian. 

As regards the claims (a) and (b), we have already observed the fact that a holomorphically 
projective transformation is analytic, and this establishes the claim (a). Conversely, a theorem 
of Matsushima \cite{matsushima} shows that every analytic vector field $\xi^{\boldsymbol a}$ 
on a K\"ahler-Einstein manifold is necessarily expressible in the form $\xi^{\boldsymbol a}=
\eta^{\boldsymbol a}+J^{{\boldsymbol a}}_{\ \boldsymbol b}\zeta^{\boldsymbol a}$, where 
$\eta^{\boldsymbol a}$ and $\zeta^{\boldsymbol a}$ are both Killing. But a Killing field on 
${\mathfrak M}$ necessarily arises from a quantum observable, from which it follows that 
$\xi^{\boldsymbol a}$ must be of the form (\ref{eq:x28}), and this establishes the claim (b). 
\vspace{0.2cm}

\noindent \textit{Remark}: The characterisation of quantum observables as eigenfunctions of the 
Laplacian need not be applicable in infinite dimension, since the trace operation is not necessarily 
valid. Nevertheless, the characterisation of a holomorphically projective transformation in the form 
$\xi^{\boldsymbol a}=\eta^{\boldsymbol a}+J^{{\boldsymbol a}}_{\ \boldsymbol b}
\zeta^{\boldsymbol a}$, where $\eta^{\boldsymbol a}$ and $\zeta^{\boldsymbol a}$ are both Killing, 
remains valid. We note that in the case of a real projective space, it has been shown in Brody \& 
Hughston \cite{BH3} that a projective transformation necessarily decomposes into a sum of a 
Killing vector and a gradient vector. The foregoing result, the mathematical content of which builds 
on Tachibana \& Ishihara \cite{ishihara}, can thus be viewed as a complex version of the 
findings in Brody \& Hughston \cite{BH3} where properties of equilibrium thermal states in classical 
statistical mechanics are investigated. 
\vspace{0.4cm}

\noindent \textbf{Discussion}. 
The present paper is focused on establishing symmetry properties of flows generated by complex 
Hamiltonians. The physical significance or implication of the result, however, remains elusive. In 
this connection it is worthwhile remarking the observation of Matveev \& Rosemann \cite{matveev} 
that complex geodesics on a quantum state space correspond to curves that lie on complex 
projective lines (which, in real terms, are just two spheres). This follows on account of the fact that 
complex projective lines are totally geodesic (geodesics on the projective line are also geodesics 
on the ambient state space); thus, along any regular curve, parallel transport of a tangent vector 
remains tangent to the line. Since a complex projective line is a two-dimensional manifold, 
its tangent plane at any point along the curve is necessarily spanned by the tangent vector and its 
rotation generated by the complex structure. 

Rephrased in a more familiar physical term, what this means is as follows. Consider a curve of 
the form 
\begin{eqnarray}
|\psiup(s)\rangle = A(s) |\etaup\rangle + B(s) |\zetaup\rangle 
\label{eq:x39}
\end{eqnarray}
that lies on a two-dimensional Hilbert space spanned by an arbitrary two distinct vectors 
$|\etaup\rangle$ 
and $|\zetaup\rangle$, where $|A(s)|^2+|B(s)|^2=1$, $A(0)=B(1)=1$. Then under the 
evolution generated by 
$\re^{{\rm i}{\hat K}t}$, where ${\hat K}$ is not necessarily Hermitian, the curve remains planer, i.e. 
at all times the curve can be expressed in the form (\ref{eq:x39}) for some time-dependent pair of 
states $|\etaup(t)\rangle$ and $|\zetaup(t)\rangle$. This observation suggests that 
the notion of a `section', i.e. a two-plane spanned by a vector and its rotation generated by the 
multiplication of the complex structure, might prove to be of importance in complex-extended 
quantum mechanics. 

Finally, a few open questions may be in order. The existence of a phase 
transition indicated above, where the characteristic behaviour of a holomorphically projective 
transformation on a complex projective space changes, appears to be unknown in the literature of 
geometry. This transition is accompanied by a geometric singularity of the following type (details of 
which will be discussed elsewhere). Suppose that the generators $H_\theta(x)$ and 
${\mathit\Gamma}_\theta(x)$ of a holomorphically projective transformation (\ref{eq:x28}) on 
${\mathfrak M}$ depend on one or a set of parameters $\theta$ (equivalently, the matrix elements 
of ${\hat K}$ depend on $\theta$), and suppose that for a given value of $\theta$, 
${\hat x}^{\boldsymbol a}(\theta)$ is a 
critical point of the flow. Then the submanifold of ${\mathfrak M}$ parameterised by $\theta$ 
exhibits curvature singularities if there are phase transitions. These transitions are typically 
accompanied by the fact that two or more of the fixed points coalesce, whereas in the generic case 
away from degeneracies there are $n$ distinct fixed points of a holomorphically projective transformation on ${\mathfrak M}$. Geometric characterisations 
and understanding of such transitions appear to be an open problem. This is of interest in physics 
because such transitions, or more generally the effects of curvature singularities on physical 
systems characterised by complex Hamiltonians, are now actively being observed in laboratory 
experiments. 

Another question concerns the investigation of the fixed-point structure of holomorphically 
projective transformations. In the case of a flow generated by a Hermitian Hamiltonian, the fixed 
points of the flow are points at which $\nabla_{\boldsymbol a}H=0$. In the case of a 
holomorphically projective transformation, if $\omega^{\boldsymbol{ab}} \nabla_{\boldsymbol{a}}
H \nabla_{\boldsymbol{b}}{\mathit\Gamma}\neq0$, then neither $\nabla_{\boldsymbol a}H$ nor 
$\nabla_{\boldsymbol a}{\mathit\Gamma}$ vanish at the fixed points. Instead, the fixed points 
are characterised by the cancellation condition: 
\begin{eqnarray}
2\omega^{\boldsymbol{ab}} \nabla_{\boldsymbol b} H = 
\varg^{\boldsymbol{ab}} \nabla_{\boldsymbol b} {\mathit\Gamma} 
\label{eq:x40}
\end{eqnarray}
at $x={\hat x}$. It will be of interest to obtain a better geometric understanding of such fixed-point 
structures.

\vspace{0.5cm}
\begin{footnotesize}
\noindent The author thanks participants of \textit{Light-matter Interaction: Focus on Novel 
Observable non-Hermitian Phenomena}, Kibbutz Ein-Gedi, Israel, April 2013, for stimulating 
discussion.
\end{footnotesize}


%
%

\end{document}